\sfcode`A=1000 \sfcode`B=1000 \sfcode`C=1000 \sfcode`D=1000
\sfcode`E=1000 \sfcode`F=1000 \sfcode`G=1000 \sfcode`H=1000
\sfcode`I=1000 \sfcode`J=1000 \sfcode`K=1000 \sfcode`L=1000
\sfcode`M=1000 \sfcode`N=1000 \sfcode`O=1000 \sfcode`P=1000
\sfcode`Q=1000 \sfcode`R=1000 \sfcode`S=1000 \sfcode`T=1000
\sfcode`U=1000 \sfcode`V=1000 \sfcode`W=1000 \sfcode`X=1000
\sfcode`Y=1000 \sfcode`Z=1000
\documentstyle[12pt,aasms4] {article}

\begin{document}

\title{The Infrared to Gamma-Ray Pulse Shape of the Crab Nebula Pulsar}

\author{Stephen S. Eikenberry and Giovanni G. Fazio}
\affil{Harvard-Smithsonian Center for Astrophysics, Cambridge, MA 02138}

\begin{abstract}

	We analyze the pulse shape of the Crab Nebula pulsar in the
near-infrared, optical, ultraviolet, X-ray, and gamma-ray bands,
including previously unpublished ROSAT HRI observations.  We show
that, in addition to the previously known trend for the fluences of
the Bridge and Peak 2 to increase with energy relative to the fluence
of Peak 1, there is a small but statistically significant trend for
both to decrease with energy relative to Peak 1 over the near-infrared
range.  We find that the phase separation between the two peaks of the
pulse profile decreases nearly continuously as a function of energy
over 7 decades of energy.  We show that the peaks' full-width
half-maxima are significantly variable over this energy range, but
without any clear pattern to the variability.  We find that the
differences between the energy dependences of the leading and trailing
edge half-width half-maxima of both peaks found by Eikenberry {\it et
al.} (1996a) also continue over 7 decades of energy.  We show that the
cusped shape of Peak 2 reverses direction between the infrared/optical
and X-ray/gamma-ray bands, while the cusped shape of Peak 1 shows weak
evidence of reversing direction between the X-ray and gamma-ray bands.
Finally, we find that many of the pulse shape parameters show maxima
or minima at energies of 0.5-1 eV, implying that an important change
in the pulsar emission is occuring near this energy.  Many of these
complex phenomena are not predicted by current pulsar emission models,
and offer new challenges for the development of such models.

\end{abstract}


\section{Introduction}

	The Crab Nebula pulsar is one of the best-studied objects in
astrophysics.  In the almost 30 years since its discovery, it has been
observed in almost every waveband, and has been particularly important
to understanding the nature of pulsars and their emission mechanisms.
One of the measurements of key interest is the the shape of the pulse
profile.  The double-peaked appearance, the sharpness of the
peaks\footnote{We refer to the 2 major features in the profile as Peak
1 and Peak 2, joined by the Bridge.  This is to avoid confusion, as
some previous authors refer to the Bridge as the ``Interpulse'', while
others use that name for Peak 2.}, and their separation by 0.4 in
phase (see Figure 1) are among the leading motivators of many of the
theories attempting to explain the pulsar emission mechanism (i.e.
Cheng et al., 1986a,b; Chiang and Romani, 1994; Romani and
Yadigoraglu, 1995; Daugherty and Harding, 1996 - hereafter CHRa,b;
CR94; YR95; DH96, respectively).  In these models of Crab Nebula
pulsar emissions, primary energy generation occurs in particle
accelerator ``gaps'' in the magnetosphere.  The accelerated particles
produce $\gamma$-rays through curvature radiation, and these
$\gamma$-rays interact with the magnetosphere in a complex manner to
produce the X-ray, ultraviolet (UV), optical, and infrared (IR)
pulsations, so that the non-radio emissions are closely linked to each
other\footnote{While some of the radio emission is thought to be
linked with the higher energy emissions, it may originate elsewhere in
the magnetosphere.  As an example, the recent work of Moffett and
Hankins (1996) shows that Peak 1 and Peak 2 are not identifiable
features over much of the radio frequency range.  For this reason, we
exclude the radio pulse shape from our analyses here.}.

	Recent work by the authors and their collaborators (Eikenberry
{\it et al.}, 1996a,b) has revealed energy dependences in the pulse
shape over the IR to UV range, some of which are challenging the
current pulsar emission models.  In particular, the half-widths of
both peaks show different energy dependences from their leading to
trailing edges, a phenomenon which is not predicted by any of the
emission models.  In this context, we are presenting new analyses of
the Crab Nebula pulsar's pulse shape from the IR to $\gamma-$ray
bands.  We are using new pulse profiles in the IR and X-ray bands with
previously unavailable signal-to-noise.  We pay particular attention
to the peak leading and trailing edge half-widths, which, as mentioned
above, are providing new challenges for the theoretical models.  Also,
in contrast to previous work on this subject (e.g.  Ramanamurthy,
1994), we treat all the pulse profiles with uniform analysis
techniques to prevent potential bias from differences in treatment.
Thus, we hope that such an analysis of the pulse shape over 7 decades
in energy will provide new insights into the nature of the pulsar
emission mechanism.

\section{Data}

\subsection{X-ray pulse profile}

	We created the X-ray pulse profile for the Crab Nebula pulsar
from a series of archival ROSAT High-Resolution Imager (HRI) sequences
originally obtained on September 2-3, 1992 (MJD 48881 and 48882).  We
downloaded 8 such sequences from the HEASARC data archive and
performed the preliminary analyses using the PROS software package.
First, we created images of the individual observations, which
revealed the pulsar as a bright point source superposed on an extended
synchrotron nebula.  Second, for each sequence we extracted all photon
events in an aperture with 40 arcsecond diameter centered on the point
source, selecting a total of $\sim954,000$ events over the
$\sim24,000$s total integration time.  Third, we corrected the photon
arrival times to the solar system barycenter, and in order to check
the X-ray timing, we epoch-folded each sequence over a range of
periods with an assumed period derivative of $\dot P = 4.20918 \times
10^{-13}$s/s and determined the period which maximized the
$\chi^2$-value for the pulse profile.  In all 8 sequences, the
measured period matched that predicted from the Jodrell Bank radio
timing ephemeris (R. Pritchard, private communication) within
statistical uncertainties (typically 0.5 to 5 nanoseconds), with $P_0
= 0.0334083300 s$ and $\dot P = 4.20918 \times 10 ^{-13}$s/s at MJD
48880.00 .  Concluding that this verifies the validity of the X-ray
timing, we then folded each sequence at the period and period
derivative given by the Jodrell Bank ephemeris into pulse profiles
with 512 phase bins, for a time resolution of $65.3 \mu$s.  Finally,
we corrected the profiles for deadtime effects, combined all 8 pulse
profiles in phase and subtracted the background, resulting in the
X-ray pulse profile shown in Figure 1.  This is the highest
signal-to-noise X-ray pulse profile of the Crab Nebula pulsar obtained
to date.

\subsection{Other pulse profiles}

	We selected the remaining pulse profiles from a variety of
sources.  For the $\gamma$-ray band, we chose the OSSE pulse profile
(Ulmer {\it et al.}, 1994) from the Compton Gamma-Ray Observatory
(CGRO) for its combination of energy coverage and high
signal-to-noise.  (We do not include data from higher energy ranges,
such as those provided by the EGRET instrument on CGRO due to the
lower signal-to-noise available at these energies).  The UV and
optical pulse profiles come from Hubble Space Telescope High-Speed
Photometer observations of the Crab Nebula pulsar (Percival {\it et
al.}, 1993), while the near-infrared pulse profiles were obtained by
the authors and their collaborators using the Solid-State
Photomultiplier (SSPM) instrument on the Multiple Mirror Telescope
(Eikenberry {\it et al.}, 1996a,b).  We present a summary of the
energy (and, where sensible, wavelength) coverage of the pulse
profiles in Table 1.

\section{Analysis - Overall pulse shape}

	The overall similarities and differences in the pulse shape
over a broad energy range from energies of $\sim1$ eV to $\sim10^6$ eV
have been known for some time, and may be clearly seen in the typical
pulse profiles we present in Figure 2.  All the pulse profiles in this
range show 2 peaks occurring at the same phase, joined by a lower (but
non-zero) flux ``Bridge''.  On the other hand, we can clearly see in
Figure 2 that the integrated flux (or fluence) of Peak 2 increases
relative to Peak 1 with energy, as does the fluence of the Bridge.  We
plot the fluence ratios of the Bridge versus Peak 1 and Peak 2 versus
Peak 1 in Figure 3, again clearly showing the overall trend for an
increase in both ratios with energy.  However, we note that there is
also a small but statistically significant trend for both fluence
ratios to decrease with energy at energies below 1 eV.  Inspection of
higher energy profiles reveals that the fluence ratios also decrease
at energies greater than $\sim 1$ MeV (Ulmer {\it et al.}, 1994).
Another feature that begins to appear with the high signal-to-noise of
the pulse profiles in Figure 2, but has been previously unnoticed, is
the shape reversal of Peak 2 - from a fast rise and slow fall in the
IR/optical range, to a slow rise and fast fall in the
X-ray/$\gamma$-ray range.  We will return to this effect and present
quantitative evidence of its existence below.

\section{Analysis - Detailed pulse shape}

	Given the overall picture of the evolution of pulse shape with
energy, we now turn to more detailed, quantitative analyses.  First,
we measure the phase separation between the 2 peaks and the full-width
half-maxima (FWHM) of the 2 peaks versus energy.  We then move on to
the analysis of the peaks' half-width half-maxima (HWHM) versus
energy, and any differences between the leading and trailing edge
HWHM.  Finally, we analyze a new shape factor for evidence of peak
shape reversal as a function of energy.

\subsection{Peak-to-peak phase separation}

	Percival {\it et al.} (1993) were among the first to show that
the phase separation between Peak 1 and Peak 2 changes with energy,
and since then many authors have measured this effect in an attempt to
understand its relation to and impact on the emission mechanism (i.e.
Ransom {\it et al.} 1994; Ramanamurthy, 1994; Ulmer {\it et al.},
1994).  However, the techniques used to measure the phase separation
have varied from author to author and from energy band to energy band,
occasionally resulting in method-dependent biases (see Eikenberry {\it
et al.}, 1996b for further discussion).

	Here, we present a measurement of the peak-to-peak phase
separation as a function of energy using the same techniques in all
energy bands.  First, we fit the central regions of each peak with a
6th-order polynomial, and take the peak position to be at the maximum
of the fit.  The phase separation is then simply the difference in
phase between the 2 maxima.  In order to estimate the uncertainties in
this technique, we perform a Monte Carlo simulation as follows.
First, we take the phase bins near the peak region and assume that the
noise in the number of counts in each bin follows a Poisson
distribution.  Next, we take the fit to the central region of the peak
and add to each bin a normally-distributed random number with a
variance corresponding to the Poisson noise for that bin.  We fit this
new profile and measure the new phase separation, and then repeat the
procedure 1000 times for each energy band and peak.  Finally, we take the
standard deviation of the simulated separations to be the $1 \sigma$
uncertainty in the measured value.  The resulting separations and
their uncertainties are presented in Figure 4 and Table 2.

	The peak-to-peak phase separation appears to be a more or less
smooth function of energy from IR to $\gamma$-ray energies.  The
separation decreases with energy over the range from 0.9 eV to $10^6$
eV.  However, there is some evidence of a turnover or break in this
trend at $E=0.7$ eV ($\lambda = 1.65 \mu$m).  While, Ulmer {\it et
al.} (1994) have shown that the peak-to-peak separation shows no
change with energy over the range $\sim50$ keV to $\sim500$ keV,
their uncertainties of $\sim0.01$ in phase are large enough to hide
the effects that we see here.

\subsection{Full-width half-maxima}

	The peak full-width half-maxima (FWHM) are easily measured
even in relatively low signal-to-noise pulse profiles (such as in
$\gamma$-ray observations), are related to the geometry of the
emission mechanism, and are thus often among the key predicted
parameters in pulsar emission models (e.g. Ho (1993)).  We measure the
FWHM of the 2 peaks from the phase difference between the points where
the profile reaches 1/2 of the peak maximum value.  In order to
determine the uncertainties in the FWHM, we perform the following
Monte Carlo simulation.  First, we fit independent third-order
polynomials to the leading and trailing edges of the peaks, and then
measure the FWHM from the points where the fit equals one-half of the
peak maximum (in all cases, the fit determination matches the original
measurement).  Second, we take the fitted counts per bin and add a
normally-distributed random number with a variance corresponding to
the Poisson noise for that bin.  We then make fits and determine the
FWHM for these new profiles.  We repeat this procedure 1000 times for
each peak, and we take the $1
\sigma$ uncertainty in the FWHM to be the standard deviation in the
Monte Carlo FWHM distribution.

	We present the resulting FWHM values and their uncertainties
in Figure 5 and Table 2.  The Peak 1 FWHM shows significant
variability from one data point to the next over much of the energy
range, but without any clear pattern.  The Peak 2 FWHM is also
variable over the IR to $\gamma$-ray energy range, and appears to show
a trend for larger FWHM at lower energies.  However, given the known
differences in the energy dependences of the HWHM values that make up
the FWHM (Eikenberry {\it et al.}, 1996a), the informational value of
the FWHM measurements is unclear.

\subsection{Half-width half-maxima}

	In previous work, we and our collaborators have found
differences in the energy dependences of the peak half-widths for the
leading and trailing edges of the IR-UV pulse profile peaks of the
Crab Nebula pulsar (Eikenberry {\it et al.}, 1996a,b).  We perform
similar analyses here to investigate whether such differences are
consistently present across this larger energy range and to determine
their form over this range.  We measure the half-widths as the phase
difference between the peak maximum and the points where the profile
drops to 1/2 of the peak maximum value.  We then perform Monte Carlo
simulations, identical to those used for the FWHM analysis, to
determine the uncertainties in the half-width half-maxima (HWHM).  We
present the resulting HWHM measurements and their uncertainties in
Figures 6 and 7 and in Table 2.

	The HWHM results show a range of interesting characteristics
in their energy dependences.  First, we note that in both peaks the
HWHM energy dependences of the leading edges differ from the trailing
edges, and the leading and trailing edge energy dependences also
differ from Peak 1 to Peak 2.  Second, we note that the difference
between leading and trailing edge HWHM energy dependence is clearly
visible in Peak 1 for the X-ray and $\gamma$-ray data points alone,
confirming that these differences do indeed persist over the entire
energy range.  Finally, we examine the particular shapes of the energy
dependences for the individual half-widths.  In Figure 6(a), we see
that the Peak 1 leading edge half-width shows a distinct maximum in
the IR at E=0.9 eV, which is either a peak in a smooth curve from 0.5
eV to $10^6$ eV, or a break between 2 curves covering this range.  The
Peak 1 trailing edge HWHM (Figure 6(b)), on the other hand, shows a
minimum in the UV (5.4 eV).  Again, this may either be part of a
smooth curve over the energy range, or evidence of a break between two
curves.  In Figure 7(a), the Peak 2 leading edge half-width shows an
apparently flat energy dependence over the IR to $\gamma$-ray range,
but with several data points showing significant deviations from the
average value.  The Peak 2 trailing edge half-width (Figure 7(b))
shows a very similar energy dependence to the Peak 1 leading edge
half-width, with a maximum at E=0.9 eV.

\subsection{Peak cusping factor}

	As noted earlier, we see in Figure 2 evidence of a reversal in
the cusped shape of Peak 2, from a short rise and long fall in the IR
profile, to a long rise and short fall in the X-ray and $\gamma$-ray
bands.  In order to quantify this behavior, we introduce a new
parameter which we call the "peak cusping factor".  The peak cusping
factor (PCF) is defined to be the logarithm of the ratio of the
leading edge half-width to the trailing edge half-width, or
\begin{equation}
\nonumber 
PCF = log_{10} \ {HWHM(lead)\over{HWHM(trail)}} .
\end{equation}
As can be seen, a change in sign in the PCF means a reversal in the
relative steepness of the leading and trailing slopes of the peak -
from a faster rise than fall to a faster fall than rise, or vice
versa.  We present the PCFs for Peaks 1 and 2 in Figure 8 and Table 2.
The Peak 1 PCF appears to be constant from 0.75 eV to 5.4 eV, with a
significantly lower value at 0.5 eV and an apparently continuous
decline from 5.4 eV to $10^6$ eV.  Note that all of the values are
positive, except for the $\gamma$-ray data point.  However, this point
lies only $\sim1.2 \sigma$ below 0, and thus fails to provide
convincing evidence of a shape reversal in Peak 1.  On the other hand,
such a reversal is clearly evident in the Peak 2 PCF (see Figures 8
and 9).  The IR-UV data points are all below 0 or consistent with
negative values, while the X-ray and $\gamma$-ray data points are
clearly positive.  Thus we see that even some of the large-scale
properties of the "canonical" Crab Nebula pulsar pulse shape, such as
a faster rise than fall in Peak 2, are in fact energy-dependent.

\section{Discussion}

\subsection{Pulsar emission models}

	We begin our discussion with a brief review of the pertinent
aspects of the pulsar emission models for the Crab Nebula pulsar, as
they provide a theoretical framework for considering the results we
present above.  One of the leading models is the 2-gap outer gap
model, proposed by Cheng, Ho, and Ruderman (CHRa,b).  In this model,
particles are accelerated in a magnetospheric vacuum gap near the
light cylinder, beaming radiation tangent to the magnetic field
direction through inverse Compton (at $\gamma$-ray energies) and
synchrotron (at lower energies) processes.  The 2 peaks of the pulse
profile arise from 2 separate gaps, each corresponding to a pulsar
magnetic pole, with the separation of 0.4 in phase resulting from the
time-of-flight delay between the 2 gaps along the observer's line of
sight.  Relativistic aberration distorts the magnetic field geometry
from that of a simple dipole in the observer's rest frame, allowing
variable time-of-flight delays from adjacent field lines, so that the
photons from separate regions may arrive at the observer at the same
time.  Thus, the sharpness of the peaks arises from a caustic in the
observed phase, where the emission from a large number of physical
regions within the gap reach the observer at the same observed phase.

	A recent variant of this model is the 1-gap outer gap model
proposed by Chiang and Romani (1994) and Romani and Yadigoraglu
(1995).  In this model, the gap extends azimuthally completely around
the pulsar, while the 2-gap model assumes that the gaps have some
limited azimuthal extent.  The emission process is fundamentally the
same as in the 2-gap model, but detailed calculations by Chiang and
Romani (1994) show that a single gap unlimited in azimuth will produce
2 caustics, and thus 2 peaks in the pulse profile.  Thus the observed
emission arises from only one gap and one magnetic pole.

	Finally, we consider a different class of model - the polar
cap (PC) gap models, specifically in the form proposed by Daugherty
and Harding (1996).  In the PC model, particles are accelerated in a
vacuum gap and emit the observed radiation tangent to the magnetic
field lines through inverse Compton and synchrotron processes (as with
the outer gap models), but in a region just above the pulsar's
magnetic pole(s).  Here, far from the light cylinder, relativistic
aberration plays a much smaller role, and the observed pulse shape
largely arises from variations in the particle distribution across the
magnetic field lines.

\subsection{Overall pulse shape}

	As noted above, our results confirm the overall trend for the
fluence of Bridge and Peak 2 to increase relative to Peak 1 with
energy.  However, the fluence ratios for both the Bridge and Peak 2
versus Peak 1 show a decrease with energy below 1 eV.  Similarly, when
Ulmer {\it et al.} (1994) subdivide the $\gamma$-ray profiles into
smaller energy bands, they find that the ratio of the Bridge + Peak 2
to Peak 1 clearly decreases over the range from a few hundred keV to a
few MeV.  Thus, while the overall trend holds, there appear to be
turnovers at both the high and low ends of the energy range.

	The overall trend for the increase with energy of the Bridge
and Peak 2 relative to Peak 1 has differing implications for each of
the models above.  In the 2-gap outer gap model, we have different
viewing angles to the 2 gaps, with a more "head-on" view of the gap
producing Peak 1.  Furthermore, the gap is subdivided into 2-3
emitting regions, each of which has different typical energies (i.e.
$\gamma$-ray, X-ray, optical/infrared), different beaming directions,
and different opening angles.  Thus different viewing angles provide
different combinations of the emissions from the various regions,
giving different overall spectra.  Therefore, the difference in
viewing angle between Peak 1 and Peak 2 may explain the observed rise
of Peak 2 with energy relative to Peak 1.  However, the 2-gap outer
gap models offer no explanation for the existence of the Bridge at
all, let alone its energy dependence.  Kamae and Sekimoto (1995) have
proposed a composite model which may remedy this situation, with the
2-gap outer gap producing Peak 1 and Peak 2, while the Bridge emission
arises from a PC gap.  In the 1-gap outer gap model, the unlimited
azimuthal extent of the gap creates Bridge emission between the 2
peaks (CR94), but no predictions of the energy dependence of the pulse
shape have been made to date.  In Polar Cap models, a hollow cone
geometry for the particle distribution can also reproduce a Crab-like
pulse shape.  While the PC models do not give explicit predictions for
the pulse shape as a function of energy, the symmetry of the current
models (DH96) does not seem to allow significant differences between
Peak 1 and Peak 2 as we observe here.

\subsection{Peak-to-peak phase separation}

	As noted above, the phase separation between Peak 1 and Peak 2
shows an apparently continuous trend to decrease with energy over much
of the IR to $\gamma$-ray energy range.  With the detailed coverage
here though, we can see that this trend is not a linear or even
constant-power-law function of energy, as has been previously
considered (e.g. Ramanamurthy, 1994).  Nevertheless, this effect again
seems to arise naturally in the 2-gap outer gap model.  In this model,
the gap is "stratified" in energy, so that the lower energy radiation
emanates from a region further from the neutron star than the higher
energy radiation.  Thus, lower energy emissions from the 2 different
gaps will have larger physical separations between them, giving
greater time-of-flight delays to the observer, and hence greater phase
separations.

	It is not clear from the current 1-gap outer gap models
whether or not the predicted phase separation should decrease with
energy.  However, energy stratification in the hollow-cone emitting
region of the PC model similar to that of the 2-gap outer gap model
may also produce a decrease in phase separation with energy.

\subsection{Peak widths (FWHM and HWHM)}

	While the peak FWHM values show significant variability with
energy over the IR to $\gamma$-ray energy range, no clear pattern for
this variability is evident.  This is not particularly surprising
given that the FWHM is composed of the leading and trailing edge HWHM,
and, as seen above, the HWHM energy dependences change from leading to
trailing edges.  Thus, we conclude only that the peak FWHM
measurements, despite their common appearance in the scientific
literature, are of dubious informational value.

	The HWHM analyses clearly show that the energy dependences of
the HWHM differ between the leading and trailing edges of the
individual peaks, confirming our earlier work in the IR-UV range
(Eikenberry {\it et al.}, 1996a).  Furthermore, the HWHM are
apparently continuous functions of energy over most, if not all, of
the energy range from $\sim0.5 - 10^7$ eV.  The possible exceptions
to this continuous behavior are the apparent breaks (or,
alternatively, maxima and minima) evident in the IR to UV ranges of
the Peak 1 leading and trailing edge HWHM and the Peak 2 trailing edge
HWHM.  This behavior may be indicative of an important change in the
pulsar emissions in the IR to UV range, though the exact nature of
this change is not currently clear.  Finally, as in Eikenberry {\it et
al.} (1996a), we note that none of these differences in the energy
dependences of the HWHM are predicted by current pulsar emission
models.

\subsection{Peak 2 shape reversal}

	The most exciting new energy-dependent characteristic of the
pulse profile that we have found is the shape reversal of Peak 2.  As
can be seen in Figure 9, Peak 2 shows a fast rise and slow fall in the
IR-optical range, but a slow rise and fast fall in the X-ray and
$\gamma$-ray range.  This work represents the first clear evidence of
such a reversal.  As with the HWHM energy dependence, this behavior is
not predicted by any current models, and offers a significant
challenge for the development of future models.  The shape reversal of
Peak 2 with energy also confirms the significant energy dependence of
the Crab Nebula pulsar's pulse shape.

\subsection{The importance of the lower energy emissions}

	Another new feature also appears in many of the analyses
presented here: the presence of maxima or minima of many of the pulse
shape parameters in the near-infrared energy range (0.5-1 eV).  The
ratio of the Bridge fluence to the Peak 1 fluence (Fig. 3a), the ratio
of the Peak 2 fluence to the Peak 1 fluence (Fig. 3b), the
peak-to-peak phase separation (Fig. 4), the Peak 1 leading edge HWHM
(Figure 6a), the Peak 2 trailing edge HWHM (Figure 7b), and the Peak 2
peak cusping factor (Fig. 8b) all show such behavior.  The fact that
so many of the pulse shape parameters show maxima or minima at 0.5-1
eV seems to imply that an important change in the pulsar emission is
occurring near this energy.

\section{Conclusion}

	We have collected pulse profiles of the Crab Nebula pulsar in
the near-infrared, optical, ultraviolet, X-ray, and $\gamma$-ray
energy bands, including previously unpublished high signal-to-noise
X-ray observations with the ROSAT High-Resolution Imager (HRI) with
$65.3 \mu$s time resolution.  For the first time, the pulse shape over
such a broad energy range has been subjected to uniform analysis
techniques, and we present the conclusions from these analyses below:

	1) The overall pulse shape consists of 2 large Peaks joined by
a non-zero Bridge of emission.  The ratios of the fluences of the
Bridge and Peak 2 relative to Peak 1 show an overall trend to increase
with energy.  While this behavior can be explained in the outer gap
emission model, the symmetry of the current polar cap model may not
allow such variations.  Furthermore, the analyses here show that the
overall trend reverses at low energies ($<1$ eV), while Ulmer {\it et
al.} (1994) have shown that the trend also reverses at high energies
($>1$ MeV).

	2) The peak-to-peak phase separation shows a non-linear trend
to decrease with energy over much of the range.  This trend is
consistent with energy stratification of the emission region in both
outer gap and polar cap emission models.

	3) The peaks' full-width half-maxima (FWHM) show significant
variability over the infrared to $\gamma$-ray energy range, but with
no definite trend or pattern.  However, such behavior matches
expectations given the known differences in the energy dependences of
the peaks' leading and trailing half-widths.

	4) The energy dependences of the peaks' half-width half-maxima
(HWHM) differ between the leading and trailing edges of the individual
peaks, confirming our earlier work in the IR-UV range (Eikenberry {\it
et al.}, 1996a).  Furthermore, the HWHM are apparently continuous
functions of energy over most of the energy range, with possible
exceptions to this continuous behavior being the apparent breaks (or,
alternatively, maxima and minima) evident in the IR to UV ranges.
These differences in the energy dependences of the HWHM are not
predicted by current pulsar emission models.

	5) The cusped shape of Peak 2 appears to reverse itself,
showing a fast rise and slow fall in the IR-optical range, but a slow
rise and fast fall in the X-ray and $\gamma$-ray range.  This work
represents the first clear evidence of such a reversal.  As with the
HWHM energy dependence, this behavior is not predicted by any current
models, and offers a significant challenge for the development of
future models.  The shape reversal of Peak 2 with energy also confirms
the significant energy dependence of the Crab Nebula pulsar's pulse
shape.

	6) Many of the pulse shape parameters show maxima or minima at
energies of 0.5-1 eV, implying that an important change in the pulsar
emission is occuring near this energy.

	7) Many of the complex phenomena we report here are not
predicted by current pulsar emission models, and offer new challenges
for the development of such models.

\acknowledgements

	We would like to thank M. Ulmer for providing the OSSE pulse
profile; J. Dolan and the HST HSP team for providing the optical and
UV pulse profiles; F. Seward and the SAO PROS support group for help
with the ROSAT data reduction; R. Pritchard for supplying the Jodrell
Bank Crab timing ephemeris in advance of publication; Team SSPM (S.
Ransom, J.  Middleditch, J.  Kristian, and C. Pennypacker) for support
of the IR research, and Rockwell International (K. Hays, M.
Stapelbroek, R.  Florence) for providing the SSPM.  S. Eikenberry is
supported by a NASA Graduate Student Researcher Program fellowship
through NASA Ames Research Center.


\begin{deluxetable} {ccc}
\tablecolumns{3}
\tablewidth{0pc}
\tablecaption{Pulse Profile Energy Coverage}
\tablehead{
\colhead{Band}		& \colhead{Energy Range}	&
\colhead{Wavelength Range}	\\ \colhead{} & 
\colhead {} & \colhead{($\mu$m)}}
\startdata
K & 0.47$-$0.56 eV & 2.00$-$2.40 \nl
H & 0.63$-$0.74 eV & 1.51$-$1.79 \nl
J & 0.81$-$0.99 eV & 1.11$-$1.39 \nl
V & 1.6$-$2.8 eV & 0.4$-$0.7 \nl
UV & 3.9$-$6.6 eV & 0.17$-$0.29 \nl
X-ray & 0.2$-$2 keV & \nodata \nl
$\gamma$-ray & 0.05$-$10 MeV & \nodata \nl

\enddata
\end{deluxetable}

 
\begin{deluxetable} {lccccccc}
\tablecolumns{8}
\tablewidth{0pc}
\tablecaption{Pulse Shape Analysis Results}
\tablehead{ \colhead{} & \multicolumn{7}{c}{Energy Band} \\
\cline{2-8} \\
\colhead{Parameter \tablenotemark{a}}	& \colhead{$\gamma$-ray} &
\colhead{X-ray}	&	\colhead{UV}		&
\colhead{V}		& \colhead{J}		&
\colhead{H}		& \colhead{K}}
\startdata
Peak-to-peak separation & 0.398 & 0.402 & 0.4045 & 0.4057 & 0.4069 &
 0.4099 & 0.4087 \nl
  & $\pm$0.003& $\pm$0.002 & $\pm$0.0015 & $\pm$0.0003 & $\pm$0.0005 & $\pm$0.0005 & $\pm$0.0003 \nl

Peak 1 FWHM & 0.041 & 0.0424 & 0.0391 & 0.04363 & 0.0489 & 0.0444 & 0.0486 \nl
  & $\pm$0.005 & $\pm$0.0010 & $\pm$0.0006 & $\pm$0.00012 & $\pm$0.0005 & $\pm$0.0010 & $\pm$0.0005 \nl

Peak 2 FWHM & 0.070 & 0.076 & 0.073 & 0.0815 & 0.091 & 0.087 & 0.086 \nl
  & $\pm$0.005 & $\pm$0.003 & $\pm$0.003 & $\pm$0.0008 & $\pm$0.003 & $\pm$0.005 & $\pm$0.004 \nl

Peak 1 HWHM (lead) & 0.016 & 0.0252 & 0.0274 & 0.0302 & 0.0339 & 0.0298 & 0.0270 \nl
  & $\pm$0.004 & $\pm$0.0018 & $\pm$0.0013 & $\pm$0.0006 & $\pm$0.0010 & $\pm$0.0013 & $\pm$0.0010 \nl

Peak 1 HWHM (trail) & 0.024 & 0.018 & 0.0117 & 0.0135 & 0.0151 & 0.0146 & 0.0215 \nl
  & $\pm$0.004 & $\pm$0.002 & $\pm$0.0012 & $\pm$0.0006 & $\pm$0.0009 & $\pm$0.0013 & $\pm$0.0010 \nl

Peak 2 HWHM (lead) & 0.043 & 0.051 & 0.037 & 0.0376 & 0.0323 & 0.039 & 0.037 \nl
  & $\pm$0.012 & $\pm$0.003 & $\pm$0.003 & $\pm$0.0007 & $\pm$0.0016 & $\pm$0.005 & $\pm$0.004 \nl

Peak 2 HWHM (trail) & 0.023 & 0.025 & 0.034 & 0.0439 & 0.059 & 0.045 & 0.048 \nl
  & $\pm$0.004 & $\pm$0.003 & $\pm$0.004 & $\pm$0.0008 & $\pm$0.002 & $\pm$0.007 & $\pm$0.005 \nl

Peak 1 PCF\tablenotemark{b} & -0.18 & 0.15 & 0.39 & 0.35 & 0.35 & 0.31 & 0.10 \nl
 & $\pm$0.13 & $\pm$0.06 & $\pm$0.05 & $\pm$0.02 & $\pm$0.03 & $\pm$0.04 & $\pm$0.03 \nl

Peak 2 PCF\tablenotemark{b} & 0.27 & 0.30 & 0.04 & -0.067 & -0.26 & -0.06 & -0.11 \nl
 & $\pm$0.12 &  $\pm$0.07 &  $\pm$0.06 &  $\pm$0.011 &  $\pm$0.03 &  $\pm$0.09 &  $\pm$0.07 \nl

\tablenotetext{a}{All parameters are in units of phase, except the PCF which is dimensionless}
\tablenotetext{b}{Peak Cusping Factor}

\enddata

\end{deluxetable}

\begin{figure}
\vspace*{130mm}
\includegraphics{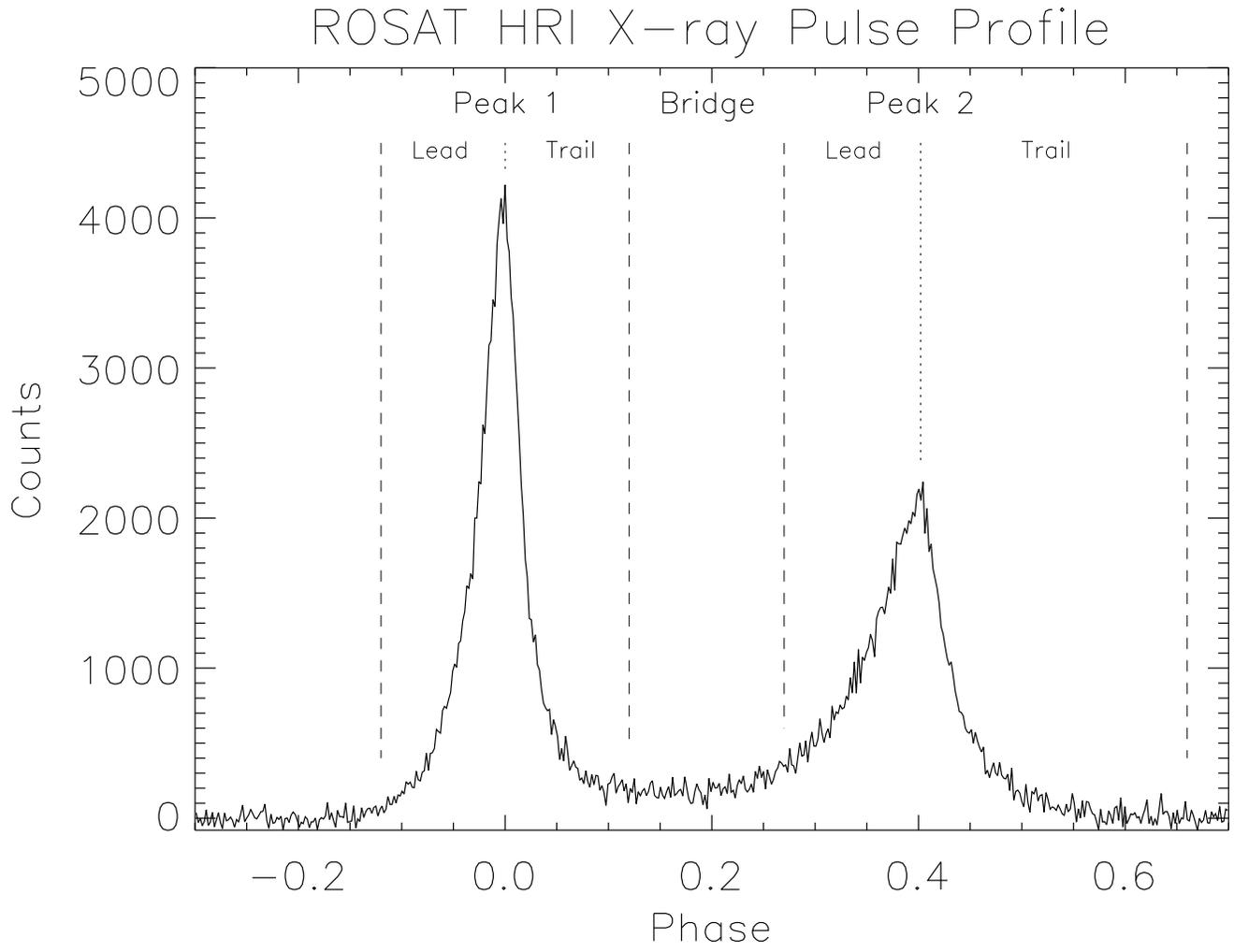}
\caption{ROSAT HRI X-ray pulse profile of the Crab Nebula pulsar with $65.3 \mu$s time resolution}
\end{figure}

\begin{figure}
\vspace*{200mm}
\includegraphics{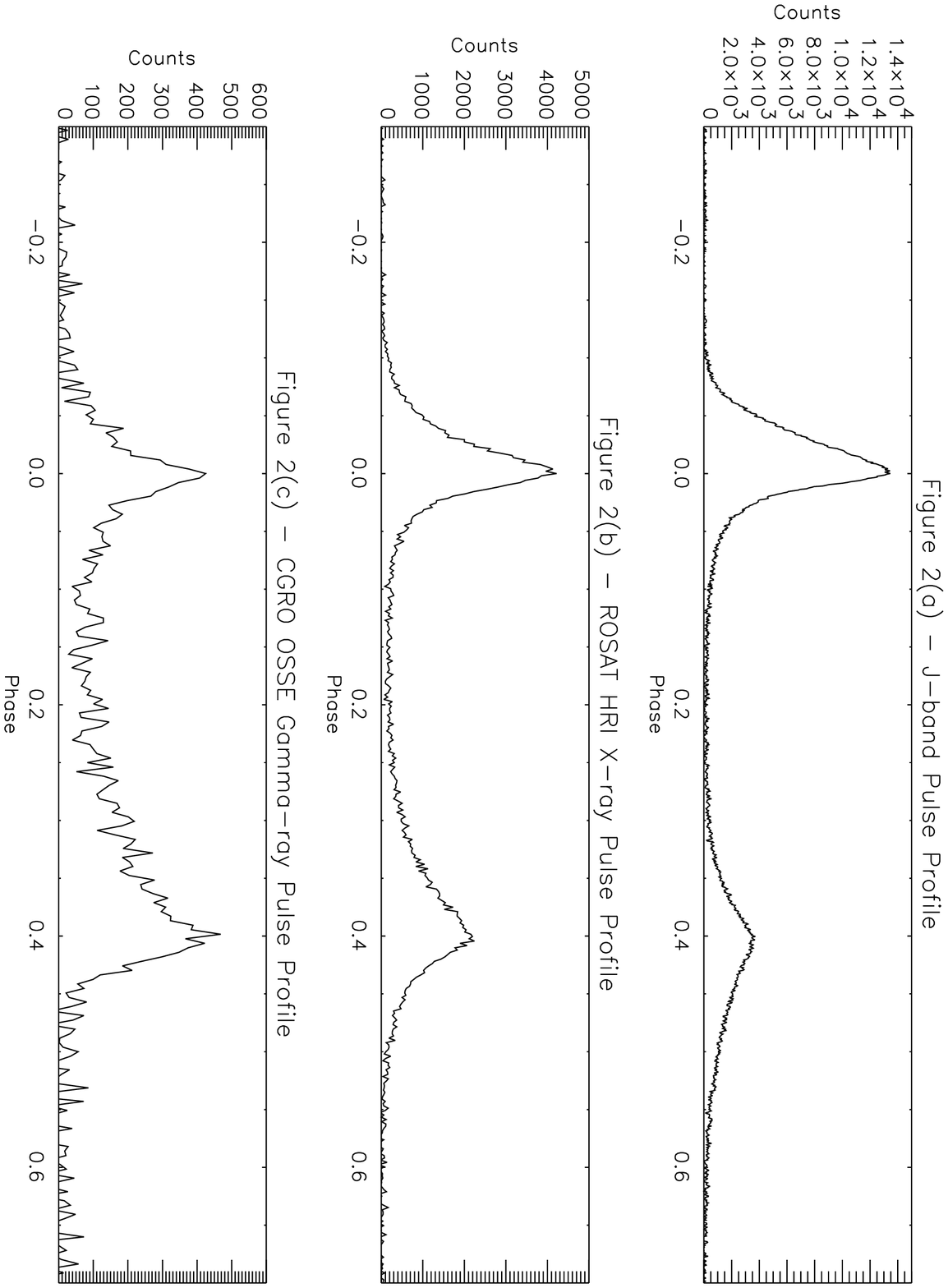}
\caption{Typical pulse profiles over the IR to $\gamma$-ray energy range: (a) J-band near-infrared profile, (b) ROSAT HRI X-ray profile, (c) CGRO OSSE $\gamma$-ray profile}
\end{figure}

\begin{figure}
\vspace*{130mm}
\includegraphics{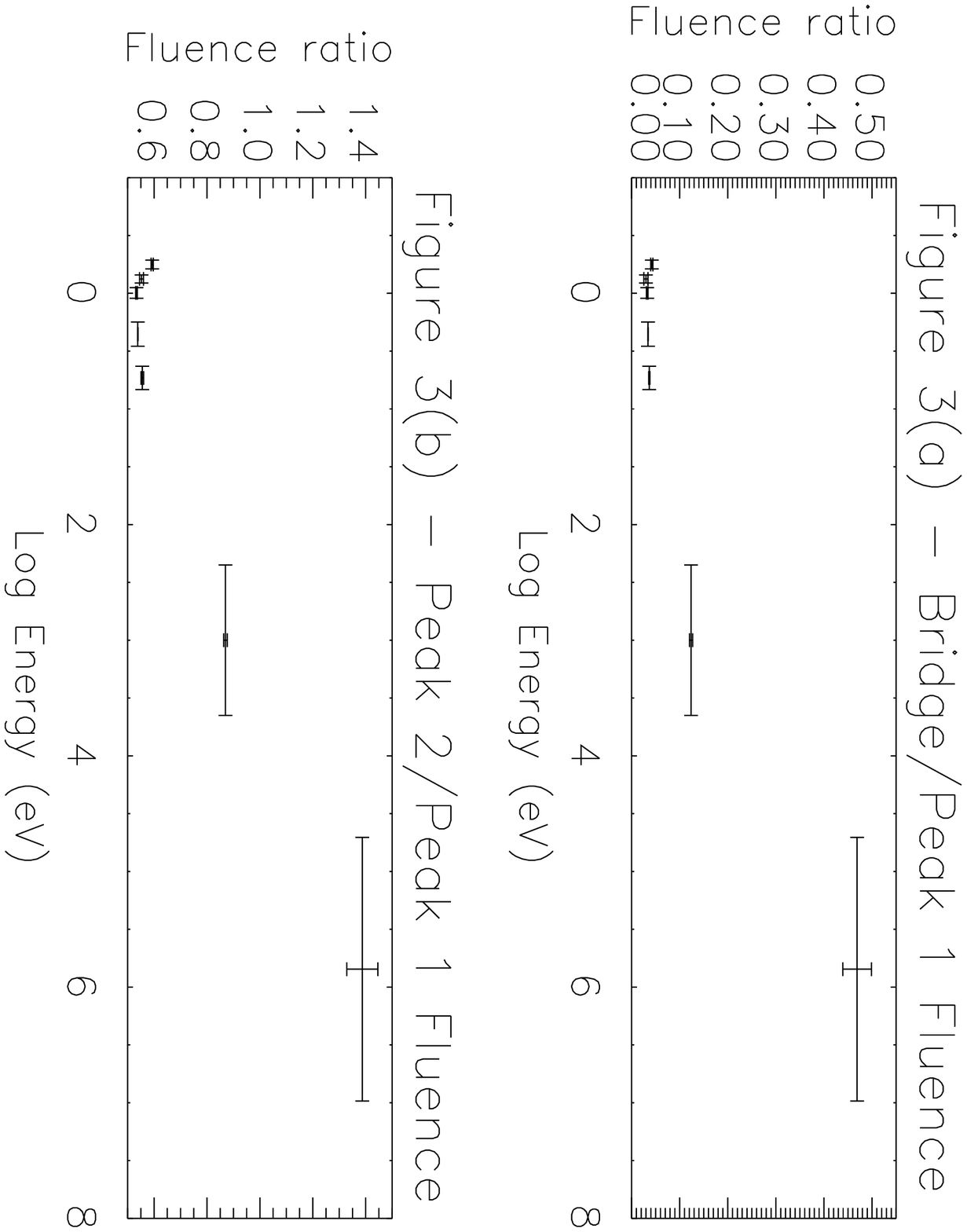}
\caption{Fluence ratios of the Bridge and Peak 2 relative to Peak 1 versus energy}
\end{figure}

\begin{figure}
\vspace*{130mm}
\includegraphics{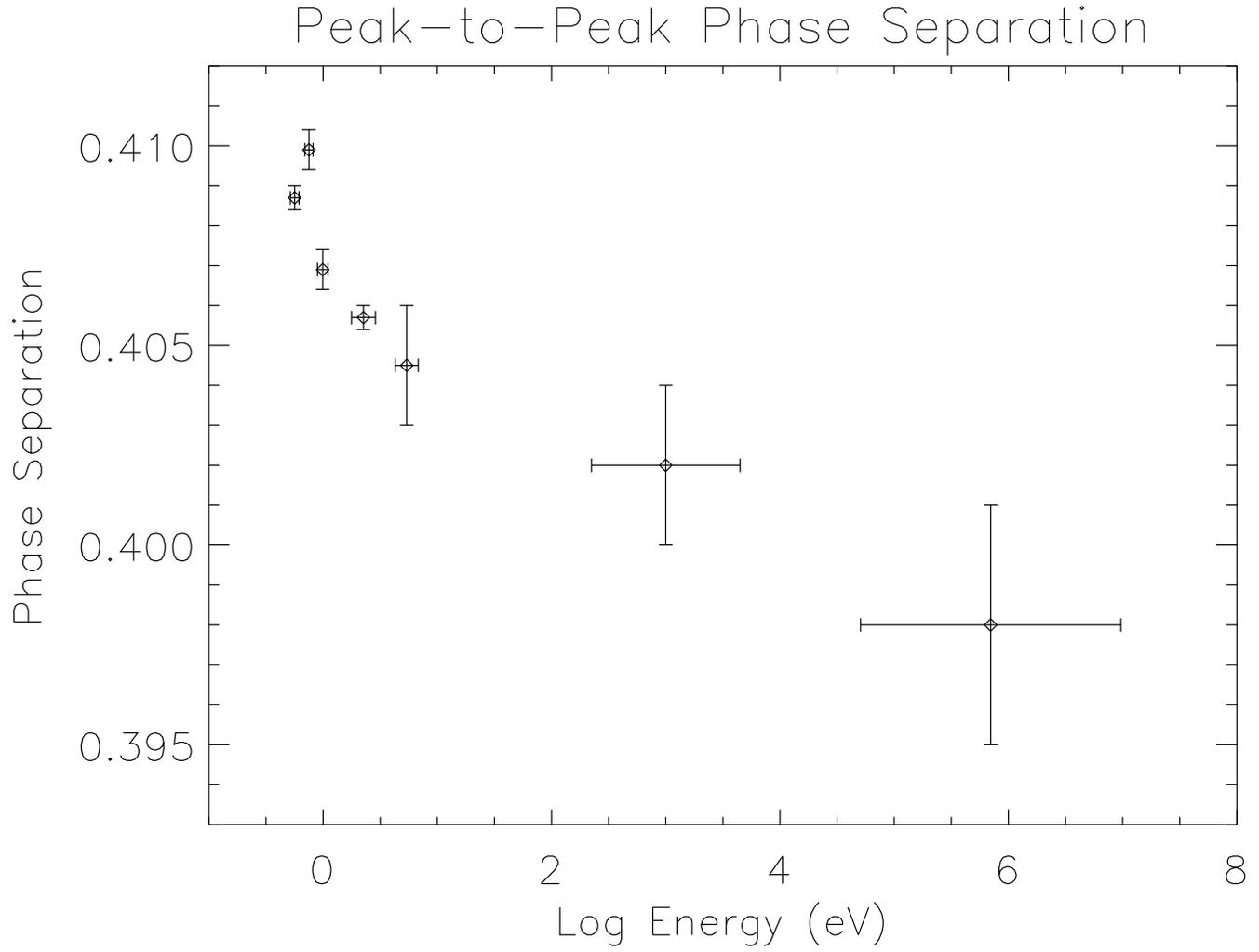}
\caption{Peak-to-peak phase separation versus energy}
\end{figure}

\begin{figure}
\vspace*{130mm}
\includegraphics{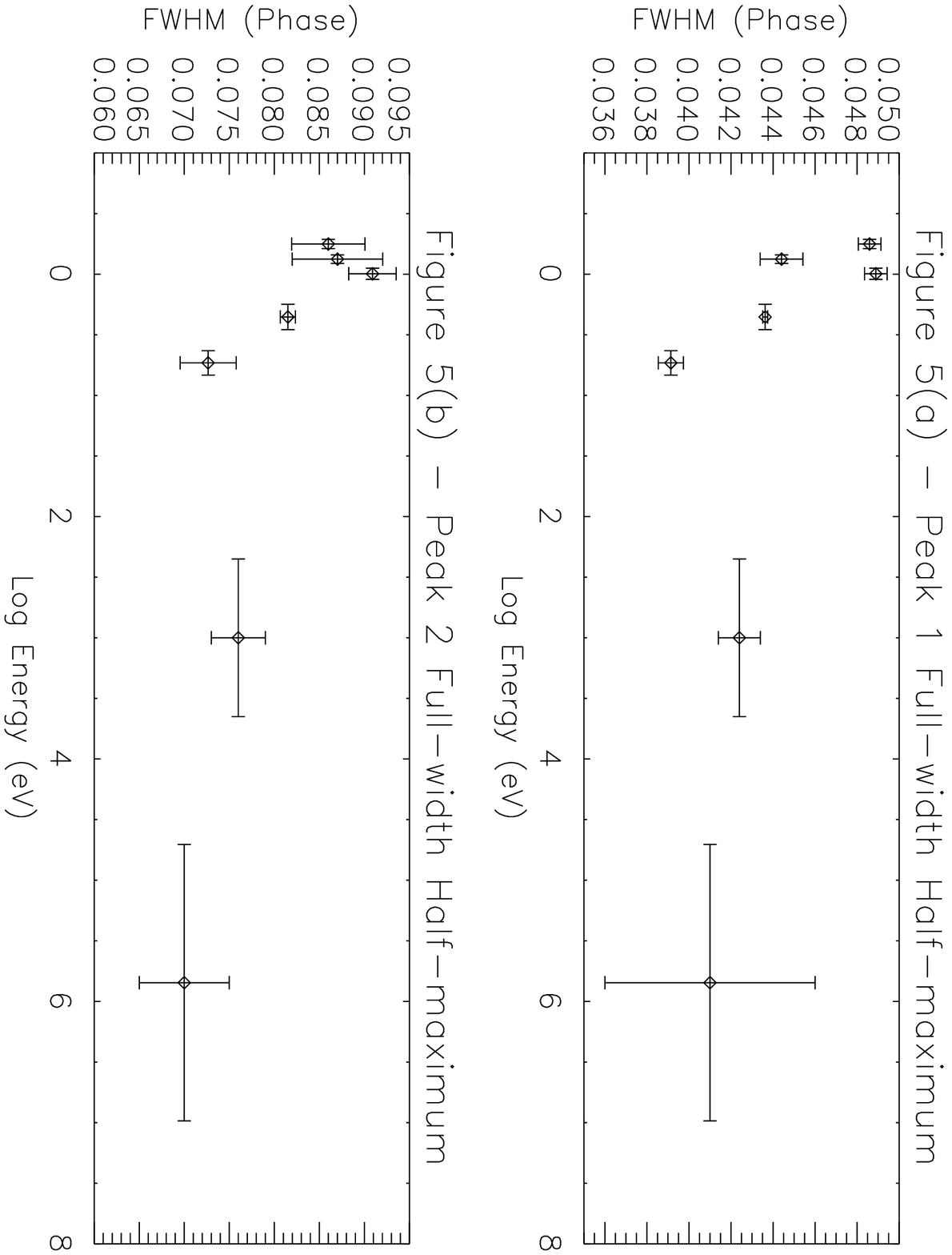}
\caption{Peak full-width half-maximum versus energy for (a) Peak 1, (b) Peak 2}
\end{figure}

\begin{figure}
\vspace*{130mm}
\includegraphics{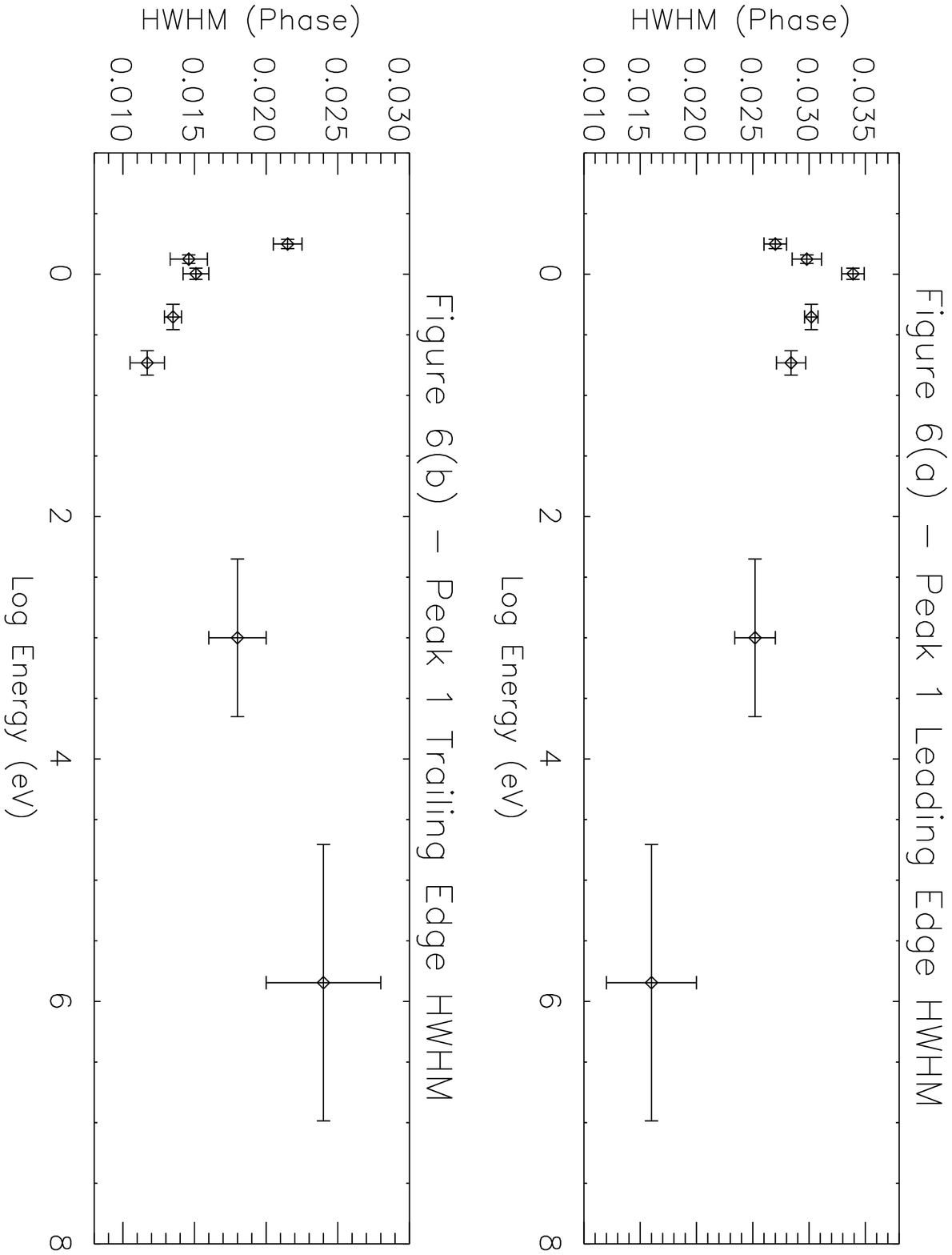}
\caption{Peak half-width half-maximum versus energy for (a) Peak 1 leading edge, (b) Peak 1 trailing edge}
\end{figure}

\begin{figure}
\vspace*{130mm}
\includegraphics{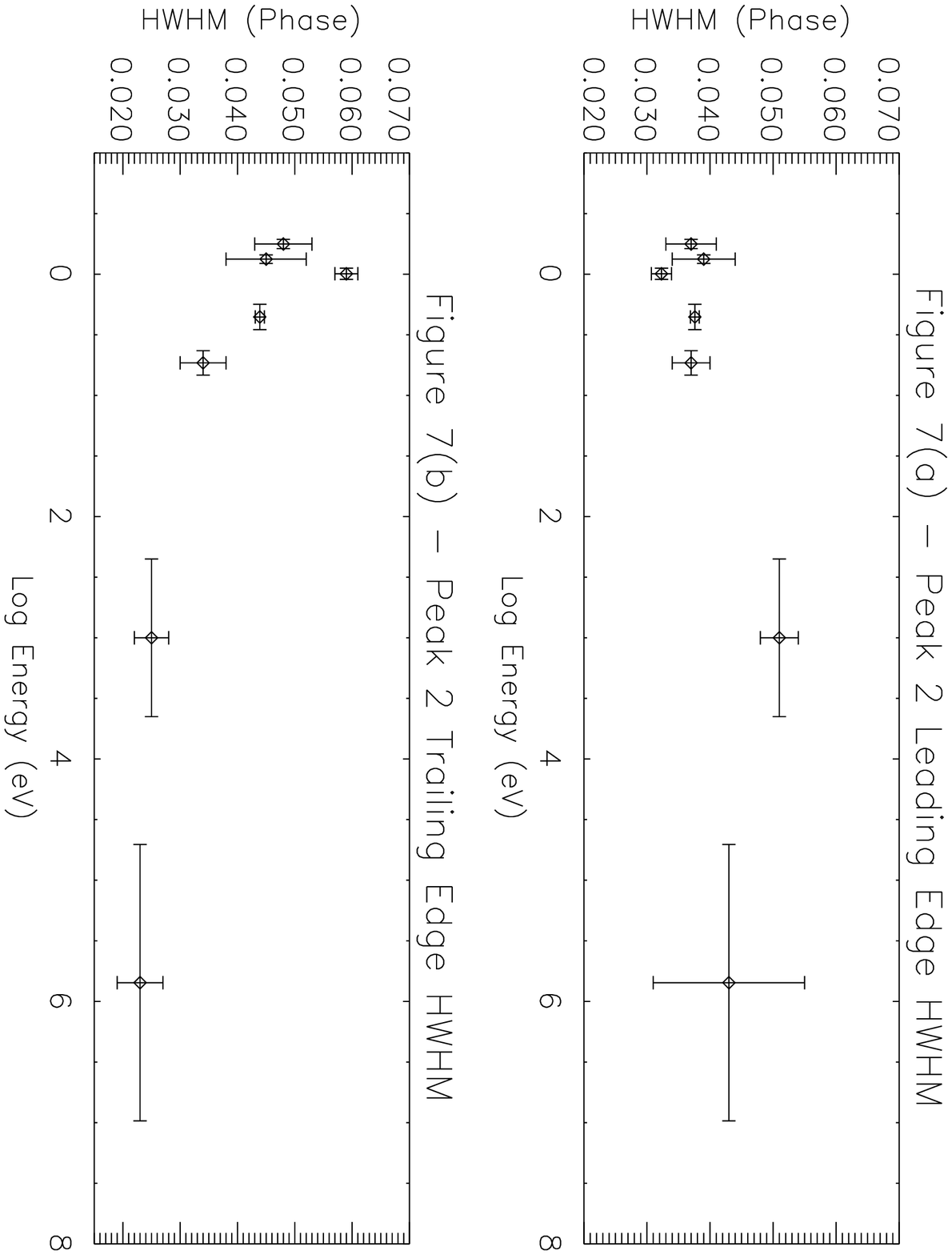}
\caption{Peak half-width half-maximum versus energy for (a) Peak 2 leading edge, (b) Peak 2 trailing edge.}
\end{figure}

\begin{figure}
\vspace*{130mm}
\includegraphics{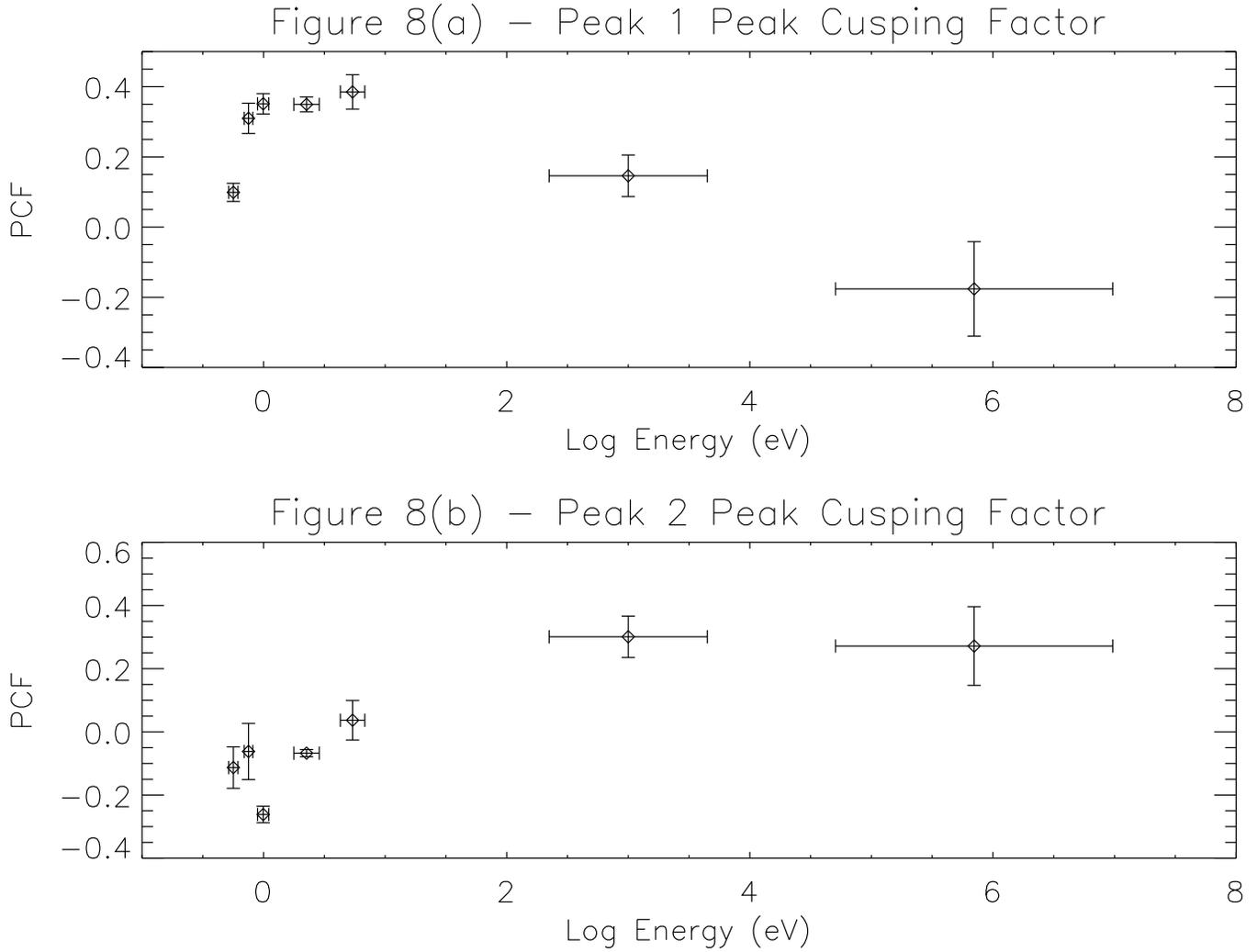}
\caption{Peak cusping factor (PCF) versus energy for (a) Peak 1, (b) Peak 2}
\end{figure}

\begin{figure}
\vspace*{200mm}
\includegraphics{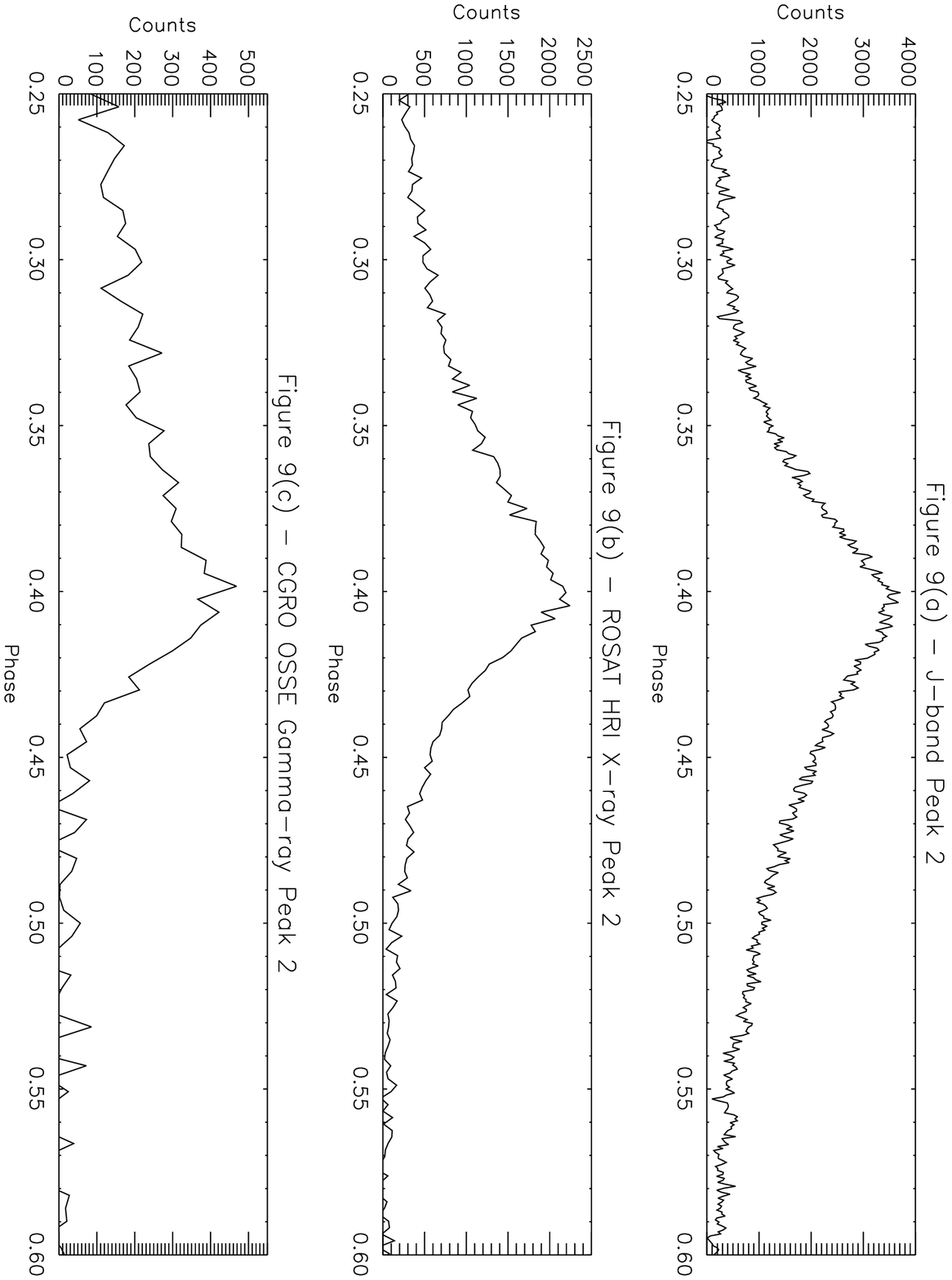}
\caption{Shape of Peak 2 for (a) J-band near-infrared profile, (b) ROSAT HRI X-ray profile, (c) CGRO OSSE $\gamma$-ray profile}
\end{figure}

\end{document}